\documentstyle[11pt,twoside,jltp]{article}

\title{Simulation of Quantum Field Theory and Gravity in
Superfluid He-3.}

\author{G.E. Volovik\address{Helsinki University of Technology,  Low Temperature
Laboratory, P.O.Box 2200, FIN-02015 HUT, Finland} \address{permanent address:
Landau Institute for Theoretical Physics, Moscow, Russia}}

\runninghead{G.E. Volovik}{Simulation of Quantum Field Theory and
Gravity.}
\begin{document}

\begin{abstract}
Superfluid phases of $3$He are quantuim liquids with the
interacting fermionic and bosonic fields. In many respects
they can simulate the interacting quantum fields in the
phyiscal vacuum. One can observe analogs of such phenomena as axial
anomaly, vacuum polarization,  zero-charge effect, fermionic
charge of the vacuum, baryogenesis,  ergoregion,  vacuum
instability, etc.  We discuss  some  topics on
example of several linear defects in $^3$He-A:  (1) disgyration, which
simulates the extremely massive cosmic string: (2) singular vortex, which is
analogous to the spinning cosmic string; and (3) continuous vortex, which motion
causes the "momentogenesis" which is the analog of baryogenesis in early
Universe. The production of the fermionic momentum by the vortex motion (the
counterpart of the electroweak baryogenesis) has been recently  measured in
Manchester experiments on rotating superfluid
$^3$He-A and $^3$He-B. To simulate the other phenomena one needs rather low
temperature and high homogeneity, which probably  can be reached under
microgravity conditions.
\end{abstract}

\maketitle

\section{Introduction. Relativistic Fermions in $^3$He-A.}

Here we demonstrate several of numerous examples of analogy between high
energy physics and superfluid  $^3$He. We consider 3 types of topologically
stable linear defects in $^3$He-A, which exhibit the properties of different
types of cosmic strings. The "gravity" and the axial anomaly in the presence
of such defects are discussed.

The quasiparticles in $^3$He-A are
chiral and massless  fermions.
Close to the gap nodes, ie   at ${\bf k}\approx \pm p_F{\hat{\bf l}}$ (
${\hat{\bf l}}$ is unit vector in the direction of the gap nodes in the
momentum space) the  energy spectrum $E({\bf k})$ of the  gapless A-phase
fermions are "relativistic"
\cite{VolovikVachaspati}:
$$E^2({\bf k}) = g^{ik}(k_i -eA_i) (k_k -eA_k)~~.\eqno(1.1)$$
Here the "vector potential" ${\bf A}$ is dynamical: ${\bf A}=k_F{\hat{\bf l}}$;
$e=\pm 1$; and the metric tensor is
$$g^{ik}= c_\perp^2 (\delta^{ik} -   l^i  l^k) +  c_\parallel^2   l^i
 l^k ~~.\eqno(1.2)$$
where $c_\perp= \Delta / p_F$ and $c_\parallel=v_F$ (with $c_\perp
\ll c_\parallel $) are  "speeds of light" propagating transverse to
${\hat{\bf l}}$
and along ${\hat{\bf l}}$ correspondingly;
$k_F$ is the Fermi momentum; $v_F=k_F/m_3$ is the Fermi velocity;
$m_3$ is the mass of $^3$He atom; $\Delta\ll k_Fv_F$ is the gap amplitude in
$^3$He-A.

In the presence of superflow with the superfluid velocity ${\bf v}_s$ the
following term is added to the energy $E({\bf k})$:
$${\bf k}\cdot {\bf v}_s \equiv ({\bf k}-e{\bf A})\cdot {\bf v}_s  +e
A_0~~,~~A_0=k_F{\hat{\bf l}}\cdot {\bf v}_s ~~.\eqno(1.3)$$
The second term corresponds to the scalar potential $A_0$ of the
"electromagnetic field", while the first one leads to the nonzero element
$g^{0i}=v_{s}^i$ of the metric tensor. As a result the Eq.(1.1) transforms to
$$ g^{\mu\nu}(k_\mu -eA_\mu) (k_\nu -eA_\nu)=0~~,\eqno(1.4)$$
with $k_\mu=({\bf k},E)$, $A_\mu=({\bf A},A_0)$ and the metric tensor
$$g^{00}=-1,~~~~g^{0i}=v_{s}^i,~~~~
g^{ik}= c_\perp^2 (\delta^{ik} -   l^i  l^k) +  c_\parallel^2   l^i
 l^k -v_{s}^iv_{s}^k,~~ \eqno(1.5)$$

\section{Disgyration as cosmic string. Conical singularity.}

The so called radial disgyration is one of the topologically stable
linear defects in
$^3$He-A. This is an axisymmetric distribution of the ${\hat{\bf l}}$ vector
$$ {\hat{\bf l}}(r,\phi)= {\hat{\bf r}} ~~,\eqno(2.1)$$
where ${\hat{\bf z}}, {\hat{\bf r}}, \hat \phi$ are unit vectors of the
cylindrical coordinate system with ${\hat{\bf z}}$ along the axis of the defect
line. The vector potential ${\bf A}=p_F{\hat{\bf r}}$ can be removed by gauge
transformation since the "magnetic" field is zero:
${\bf B}=\vec\nabla\times{\bf A}=0$. Thus the radial disgyration provides only
the "gravity" field, acting on the $^3$He-A fermions, with the metric tensor
$$g^{00}=-1,~~~~g^{0i}=0,~~~~
g^{ik}= c_\perp^2 ({\hat{\bf z}}^i{\hat{\bf z}}^k +  \hat \phi^i  \hat
\phi^k) +
c_\parallel^2   {\hat{\bf r}}^i
 {\hat{\bf r}}^k  .~~ \eqno(2.2)$$
The interval corresponding to this metric
$$ds^2=-dt^2 + {1\over  c_\perp^2} dz^2 + {1\over  c_\parallel^2}
\left(dr^2 +  { c_\parallel^2 \over  c_\perp^2  } r^2 d\phi^2\right) ~~,
\eqno(2.3)$$
has a conical singularity (if
$c_\perp\neq c_\parallel$) with curvarture being concentrated at the axis of
disgyration ($r=0$) \cite{SokolovStarobinsky,Banados}.  The space is
flat everywhere outside the axis but the length of circumference of radius $r$
around the axis is
$2\pi r  (c_\parallel/  c_\perp)$. Since
$c_\parallel \gg  c_\perp$, this is analogous to rather unusual cosmic string
with a very big positive or negative mass $M$ per unit length:
$$ 4GM=1\pm{ c_\parallel  \over  c_\perp   }  ~~,~~  4G\vert M\vert
\gg 1~~.
\eqno(2.4)$$
Since the linear mass $M=1/4G$ corresponds to the chain of the point
masses $m$ with the distance between the neighbouring masses equal
twice the Schwarzschild radius ($r_g=2Gm$: $M=m/2r_g=1/4G$) the case
$\vert M \vert >1/4G$ corresponds to the chain of the overlapping black holes.

\section{ Symmetric vortex and spinning string.}

The most symmetric vortex in $^3$He-A can be realized in thin films where the
vector
${\hat{\bf l}}$ is fixed along the normal to the film, while the superfluid
velocity
is circulating around the vortex axis:
$$ {\hat{\bf l}}= {\hat{\bf z}}~~,~~ {\bf v}_s={\hbar\over 2m_3
r}\hat \phi~~.\eqno(3.1)$$
The "magnetic" field and the scalar potential are absent (${\bf
B}=k_F\vec\nabla\times{\hat{\bf l}}=0$, $A_0=0$). One has again
only the gravity field,  now with the metric tensor
$$g^{00}=-1,~~~~g^{0i}={\hbar\over 2m_3
r}\hat \phi^i,~~~~
g^{ik}= c_\parallel^2   z^i
 z^k+ c_\perp^2 (x^i  x^k+y^i  y^k)    -v_{s}^iv_{s}^k.~~ \eqno(3.2)$$
The corresponding interval is \cite{ThreeForces}
$$
ds^2=-\left(1-{v_s^2(r)\over c_\perp^2}\right)\left ( dt +{\hbar d\phi\over 2
m_3( c_\perp^2-v_s^2(r))}  \right )^2 + {dz^2\over c_\parallel^2}+
{dr^2\over c_\perp^2}  +{  r^2d\phi^2\over
 c_\perp^2- v_s^2(r) } ~~.
\eqno(3.3)
$$
There are two important properties of this interval:

(1) There is a region, where
the velocity field
$v_s(r)$ exceeds the transverse speed of light, $ c_\perp$. This is the
`ergoregion': the vacuum in the ergoregion is unstable towards creation of pairs
of particles.

(2) Far from the vortex axis, where $v_s(r)$ is small and can be
neglected:
$$
ds^2=- \left( dt + {d\phi\over \omega}\right)^2 +{1\over c_\parallel^2}dz^2+
{1\over c_\perp^2}(dr^2 +r^2d\phi^2)~~,
\eqno(3.4)
$$
where the angular velocity
$$\omega={ 2 m_3 c_\perp^2  \over \hbar}~~.
\eqno(3.5)$$
This metric corresponds to that outside
the so called spinning cosmic string
\cite{Mazur}, which has the angular momentum in the core. In our case this is
unusual spinning string  with the angular momentum
$J=
\hbar/ (8 m_3 G)$ per unit length ($G$ is the gravitational constant), but with
zero mass.

The connection between the time and the angle in Eq.(3.4) suggests \cite{Mazur}
that the energy
$E$ and angular momentum $J$ of fermions on the background of this
spinning string are related as
$$E =J\omega~~.
\eqno(3.6)
$$
The spectrum of bound states in the core of this vortex was calculated by Kopnin
\cite{Kopnin}. He found that the factor $\omega$ in  Eq.(3.6) is of  the same
order, though is not equal to that in Eq.(3.5). What is more important,
the calculated $\omega$ appeared to be independent of the momentum $k_z$ along
the vortex axis in a complete agreement with Eq.(3.5). This is distinct from the
spectrum of bound states in the core of vortices in other systems: In
$s$-wave superconductors and in $^3$He-B there is an essential dependence of
$\omega(k_z)$ on $k_z$.\cite{Caroli}

\section{ ATC vortex and baryogenesis.}

The continuous vortex, discussed by Chechetkin \cite{Chechetkin} and
Anderson and Toulouse\cite{AT} (ATC vortex), has the following distribution of
${\hat{\bf l}}$ and ${\bf v}_s$ fields:
$$
{\hat{\bf l}}(r,\phi)={\hat{\bf z}} \cos\eta(r) + {\hat{\bf
r}} \sin\eta(r)~,~{\bf v}_s(r,\phi)=-{\hbar\over 2 m_3 r}[1+\cos\eta(r)]{\hat
{\bf \phi}}~~,
\eqno(4.1)
$$
where $\eta(r)$ changes from $\eta(0)=\pi$ to $\eta(\infty)=0$ in the so called
soft core of the vortex. The stationary vortex generates the
``magnetic'' field and when the vortex moves with a
constant velocity ${\bf v}_L$ it also generates
the ``electric''  field, since ${\bf A}$  depends  on
${\bf r}-{\bf v}_Lt$:
$${\bf B}=k_F\vec \nabla \times {\hat {\bf l}}~~,~~{\bf E}=\partial_t {\bf
A}=-k_F({\bf v}_L\cdot
\vec\nabla){\hat {\bf l}} ~~. \eqno(4.2)
$$
The quantum vacuum with massles
chiral fermions exhibits the axial anomaly:  the presence of the
electric and magnetic fields leads to the production of left particles from the
vacuum at the rate\cite{Adler1969,BellJackiw1969}
$$
\dot{n}=\partial_\mu j^\mu ={e^2 \over {4\pi^2}} {\bf E} \cdot {\bf
B}~~. \eqno(4.3)
$$

In $^3$He-A there are two species of fermions,  left and right, with $e=\pm 1$.
The left quasiparticle carries the  linear momentum $p_F{\hat {\bf l}}$, while
the righthanded quasiparticle carries the opposite momentum $-p_F{\hat {\bf
l}}$. This asymmetry between left and right gives the net product of the
fermionic linear momentum
${\bf P}$ in the time-dependent texture:
$$
\partial_t {\bf P}=
2\int d^3r~\dot{n}~ p_F\hat {\bf l}=
\hbar{k_F^3\over {2\pi^2}}\int d^3r~  \hat {\bf l} ~(\partial_t \hat {\bf l}
\cdot (\vec \nabla \times \hat {\bf l} \, \, ))~~. \eqno(4.4)
$$

Integration of the anomalous momentum transfer in
Eq.(4.4) over the cross-section of the soft core of the moving ATC vortex gives
the rate of the momentum transfer between the condensate (vacuum) and the heat
bath (matter),  mediated by the moving vortex \cite{Volovik1992}:
 $$ \partial_t {\bf P}=-\pi \hbar N_v  C_0{\hat {\bf z}}  \times {\bf
v}_L ~~.
\eqno(4.5)
$$
Here $N_v$ is the winding number of the vortex ($N_v=-2$ in the case of
Eq.(4.1)) and $C_0=  k_F^3/3\pi^2$.
This "momentogenesis" from the vacuum gives an extra nondissipative force acting
on the moving continuous vortex.
This result, derived  for the ATC vortex from the axial anomaly equation
(4.3), was confirmed in the microscopic theory, which took into accout the
discreteness of the quasiparticle spectrum in the soft core \cite{Kopnin1993}.
This was also confirmed in experiments on vortex dynamics in $^3$He-A
\cite{BevanNature}, where it was found that the extra force on the vortex
nearly cancels the conventional Magnus force.

In the Weinberg-Salam model the similar asymmetry between left and right
leads to
the "baryogenesis" -- production of the baryonic charge in the presence of the
$SU(2)$ and $U(1)$ fields. Such fields  can be generated in
the core of the topological defects  (monopoles, domain walls, sphalerons
and electroweak cosmic strings) evolving in the expanding Universe
\cite{Dolgov,Vilenkin,HindmarshKibble,Turok,tvgf,jgtv}. Experiments in $^3$He-A
and also in $^3$He-B\cite{BevanNature}, where the "momentogenesis" due to the
axial anomaly in the singular core has been measured in a broad temperature
range, support ideas on electroweak baryogenesis in early Universe.

\section{Conclusion. 3He droplets in microgravity.}

We discussed here only a very small fraction of the analogies which can
simulate numerous phenomena in particle physics and gravitation.

In superfluid
$^3$He it is very often that the container walls  prevent the conducting of the
"pure" experiments. Such nonsuperfluid environment should be removed since the
Universe, according to our present knowledge, has no external environment.  One
example: in the process of nucleation of quantized vortices (the phenomenon
which is now believed to be   important in the cosmological models
\cite{Ruutu}) the surface roughness of container dominates over the intrinsic
nucleation. But only the latter is interesting for the cosmological
community.
The absence of gravity gives us a chance to produce the free droplet
without the container boundaries and thus to deal with the pure intrinsic
nucleation. The rotation of a free droplet  $^3$He-A liquid can be made by
utilizing the unique magnetic properties of superfluid 3He: one can apply the
rotating magnetic field.

Another interesting   problem  related to the droplets of  $^3$He is the
dependence of its superfluid properties on the size of droplet. The extreme
case --  the cluster with a small  number $N$ of   $^3$He atoms  --
corresponds to the other finite system -- atomic nucleus which also represents
the cluster of fermions (protons and neutrons).   These clusters have very
similar properties, determined by their fermionic quantum statistics and
interaction: shell structure, magic numbers, single and collective
excitations, rotational degrees of freedom, fission and
fusion,  deformed and superdeformed states of
clusters with large angular momentum, the superfluid (pair-correlated)
properties of nuclei which evolve with increasing $N$, etc. The advantage of
$^3$He clusters, that it is relatively easy to study them using the NMR
technique.

I thank T. Jacobson, P. Mazur and K. Rama for discussions.

\end{document}